\documentclass[a4paper]{article}
\usepackage[english]{babel}
\usepackage[utf8]{inputenc}
\usepackage{amsmath}
\usepackage{graphicx}
\usepackage[colorinlistoftodos]{todonotes}
\usepackage{fullpage}
\def \be {\begin{equation}}
\def \ee {\end{equation}}
\def \ba {\begin{array}}
\def \ea {\end{array}}
\def \bea{\begin{eqnarray}}
\def \eea{\end{eqnarray}}
\def \nn {\nonumber}
\def \Tr {{\textrm{Tr}}}
\def \tr {{\textrm{tr}}}

\def \b {\beta}
\def \g {\gamma}

\def \d {\delta}

\def \r {\rho}

\def \hs {\hspace}

\title{\textbf{Holographic R\'enyi Entropy of Single Interval on Torus: with $\cal W$ symmetry}}
\author{
Bin Chen$^{1,2,3}$\footnote{bchen01@pku.edu.cn}\,
Jie-qiang Wu$^{1}$\footnote{jieqiangwu@pku.edu.cn}\,
and Ze-chuan Zheng$^{1}$\footnote{1200011602@pku.edu.cn}
}
\date{}

\begin{document}

\maketitle

\begin{center}
{\it
$^{1}$Department of Physics and State Key Laboratory of Nuclear Physics and Technology, Peking University, Beijing 100871, P.R.\! China\\
\vspace{2mm}
$^{2}$Collaborative Innovation Center of Quantum Matter, 5 Yiheyuan Rd, \\Beijing 100871, P.~R.~China\\
$^{3}$Center for High Energy Physics, Peking University, 5 Yiheyuan Rd, \\Beijing 100871, P.~R.~China
}
\vspace{10mm}
\end{center}

\begin{abstract}
In this paper, we extend our study of the holographic R\'enyi entropy of single interval on a torus to the CFT with ${\cal W}$ symmetry. To read the possible corrections from ${\cal W}$ fields clearly, we compute the thermal correction to the R\'enyi entropy up to the order $e^{-8\pi /TL}$. In the field theory, this requires us to compute the contribution from all the level 4 states, from both the vacuum module and ${\cal W}$ primaries. On the gravity side, after computing the eigenvalue of single Schottky generators to the order $e^{-4 \pi/TL}$, we obtain the 1-loop quantum correction to the holographic R\'enyi entropy. We find complete agreement between the bulk and boundary theories.
\end{abstract}

\section{Introduction}

The holographic computation of the entanglement entropy provides a new window to study the AdS/CFT correspondence. In \cite{Ryu:2006bv,Ryu:2006ef}, it was proposed by Ryu and Takayanagi that the entanglement entropy of a subsystem $A$ in a conformal field theory could be holographically computed by the area of a minimal surface in the dual AdS spacetime, which is homogeneous to $A$
 \be S_{EE}= \frac{\mbox{Area}(\Sigma_A)}{4G}.  \label{RT}\ee
The resulting entanglement entropy is called holographic entanglement entropy.
  It could be understood as a kind of generalized gravitational entropy \cite{Lewkowycz:2013nqa}  \cite{Fursaev:2006ih}, in the framework of the Euclidean gravity. Since its proposal, the holographic entanglement entropy has been studied from various points of view( see the nice reviews \cite{Nishioka:2009un,Takayanagi:2012kg}). It not only provides an effective way to compute the entanglement entropy in quantum field theory, but also sheds new light on the holographic nature of quantum gravity, at the semiclassical level.

The study of R\'enyi entropy in the context of AdS$_3$/CFT$_2$ correspondence is of particular importance. In this case, the quantum gravity AdS$_3$ spacetime is dual to a two-dimensional(2D) conformal field theory (CFT) with a central charge \cite{Brown:1986nw}
\be c=\frac{3l}{2G}, \ee
 where $G$ is the 3D gravity coupling constant and $l$ is the AdS radius. Though the explicit construction of the dual CFT is not completely clear, it is expected to have a sparse light spectrum \cite{Hartman:2013mia} \cite{Hartman:2014oaa}. Unlike the higher dimensional AdS/CFT correspondence, there exists a nice limit in AdS$_3$/CFT$_2$, under which the computation on both  sides are simplified. This is the large central charge limit.  In this limit, the conformal blocks in the CFT are simplified and the contribution to the R\'enyi entropy is dominated by the vacuum Verma module\cite{Hartman:2013mia}. On the other hand, the bulk gravity becomes weakly coupled and the on-shell semi-classical action of the  gravitational configuration ending on the boundary Riemann surface gives the holographic R\'enyi entropy\cite{Headrick:2010zt,Faulkner:2013yia}. For the multi-interval case, the studies in \cite{Hartman:2013mia,Faulkner:2013yia} show that the RT formula (\ref{RT}) is exact in the large central charge limit. Actually, it is the leading contribution in the large central charge limit which gives the RT area law. Moreover, the next-leading contribution in the field theory, which is of order one, should correspond to the 1-loop quantum correction to the classical bulk action \cite{Headrick:2010zt,Barrella:2013wja}. The 1-loop quantum correction to the holographic R\'enyi entropy has been studied carefully in the case of double intervals with a small cross ratio\cite{Chen:2013kpa,Chen:2013dxa,Beccaria:2014lqa} and the case of single interval on a torus\cite{Chen:2014unl}. The remarkably good agreements between  field theory and gravity have been obtained.

In this work, we extend our study on the  R\'enyi entropy of single interval
on a torus to the CFT with $\cal W$ symmetry. In this case, the dual gravity could be a higher spin (HS) AdS$_3$ gravity theory\cite{Campoleoni:2010zq,Henneaux:2010xg}. Therefore the study of the  R\'enyi entropy in the CFT with $\cal W$ symmetry opens a new window to  understand the HS/CFT correspondence. In \cite{Chen:2013dxa,Perlmutter:2013paa}, the double-interval on a complex plane has been discussed carefully. On the field theory side, the CFT is still in the vacuum, and the vacuum module dominates the leading contribution. However, the module from $\cal W$ primary field does contribute at the next-leading order.
On the bulk side, the semi-classical gravitational configuration is the same as the one in pure gravity without higher spin fields, but the fluctuations should include the higher spin ones. In other words, the contributions from the higher spin fields to the R\'enyi entropy only appear at the quantum level. For the case of single interval on a torus, the situation is different. Now the CFT is defined on a circle and at a finite temperature. As a result, there must be thermal correction and finite size correction to the entropy. In this work, we focus on the finite temperature case without turning on the higher spin chemical potential so that the dual gravitational configurations carry no higher spin charge. Therefore, we are allowed to use the knowledge in \cite{Barrella:2013wja,Chen:2014unl} to read the classical part of holographic R\'enyi entropy. However, to read the higher spin contributions to the quantum part, we have to develop the perturbative computation of the eigenvalues of the primitive conjugate class elements to a higher order. On the field theory side, we work on the CFT at the low temperature such that we can expand the thermal density matrix level by level. In order to read the contribution from the $\cal W$ primary field clearly, we need to read the next-leading contributions of the level $4$ operators. Finally, we find perfect agreements between field theory and gravity computations\footnote{For other studies on the entanglement entropy in a CFT with $\cal W$ symmetry, see \cite{Ammon:2013hba,deBoer:2013vca,Datta:2014ska,Datta:2014uxa,Long:2014oxa,Castro:2014mza,Datta:2014zpa,deBoer:2014sna}.}.

In the next two sections, we present the computations in CFT and gravity respectively. We conclude with some discussion. In the appendix, we collect the detailed results on the level 4 contributions.

\section{CFT computation}

In quantum field theory, the entanglement entropy is defined with respect to the spacelike submanifold $A$ at a fixed time. For a density matrix of the whole system, the reduced density matrix for $A$ is obtained by tracing out the states of its complement
\be \rho_A=\tr_B\rho. \ee
The entanglement entropy is defined to be the von Neumann entropy of the reduced density matrix
\be
S_{EE}=-\Tr_A \r_A\log \r_A.
\ee
It could be read from
 the Renyi entropy defined as
\be\label{Renyi} S_n=\frac{1}{1-n}\log\Tr({\rho_A}^n), \ee
by considering the $n\to 1$ limit
\be
S_{EE}=\lim_{n\to 1} S_n.
\ee

In this section, let us compute the single interval entangle entropy and Renyi entropy on a circle at low temperature. We set the circle length to be $L$, the interval length to be $l$, and the temperature to be $T=\frac{1}{\b}$. In the field theory, we can compute the  Renyi entropy and entanglement entropy via the replica trick.  At a finite temperature, the density matrix becomes thermal
\be \rho=\frac{e^{-\b H}}{\Tr e^{-\beta H}}=\frac{1}{\Tr e^{-\beta H}}\sum\mid\phi\rangle\langle\phi\mid e^{-\beta E_{\phi}} \ee
where the summation is over all the excitations in the theory.
On a cylinder the energy spectrum is read by
\be H=\frac{2\pi}{L}(L_0+\widetilde{L_0}-\frac{c}{12}). \ee
Thus we can expand the thermal density matrix according to the level. In this work, we are satisfied to consider the excitations up to level four at a low temperature.
The essential point is that expanding the thermal density matrix is equivalent to insert the a complete set of state bases along the thermal cycle such that the torus becomes a cylinder. By a uniformization conformal map and the state-operator correspondence, the computation is recast into the sum of multi-point functions on the $n$-sheeted cylinder\cite{Cardy:2014jwa}. 

First let us consider the vacuum module.  The level two states and their corresponding vertex operators are respectively
\bea
	|2\rangle=\sqrt{\frac{2}{c}}L_{-2}\mid 0\rangle&\rightarrow&\sqrt{\frac{2}{c}}T(u),\nn\\
\mid \widetilde{2}\rangle=\sqrt{\frac{2}{c}}\widetilde{L_{-2}}\mid 0\rangle&\rightarrow&\sqrt{\frac{2}{c}}\widetilde{T}(\bar{u}).
	\eea
The level three states and their corresponding vertex operators are respectively
\bea
	|3,1\rangle=\sqrt{\frac{1}{2c}}L_{-3}\mid 0\rangle &\rightarrow&\sqrt{\frac{1}{2c}}\partial T(u), \nn\\
\mid \widetilde{3},1\rangle=\sqrt{\frac{1}{2c}}\widetilde{L_{-3}}\mid 0\rangle&\rightarrow&\sqrt{\frac{1}{2c}}\partial \widetilde{T}(\bar{u}).
	\eea
And the level four states and their corresponding vertex operators are respectively
\bea
	|4,1\rangle&=&\sqrt{\frac{1}{5c}}L_{-4}\mid 0\rangle\rightarrow  \sqrt{\frac{1}{5c}} \frac{1}{2}\partial^2 T(u),\nn\\
\mid\widetilde{4,1}\rangle&=&\sqrt{\frac{1}{5c}}\widetilde{L_{-4}}\mid 0\rangle\rightarrow \sqrt{\frac{1}{5c}} \frac{1}{2}\partial^2 \widetilde{T}(\bar{u}),\nn\\
|4,2\rangle&=&(\frac{c^2}{2}+\frac{11}{5}c)^{-\frac{1}{2}}(L_{-2}L_{-2}-\frac{3}{5}L_{-4}) \mid 0\rangle \nn\\
 &\rightarrow & \left(\frac{c^2}{2}+\frac{11}{5} c\right)^{-1/2}\left(:T(u)^2:-\frac{3}{10}\partial^2 T(u)\right),\nn\\
 \mid \widetilde{4,2}\rangle&=&(\frac{c^2}{2}+\frac{11}{5}c)^{-\frac{1}{2}}(\widetilde{L_{-2}}\widetilde{L_{-2}}-\frac{3}{5}\widetilde{L_{-4}}) \mid 0\rangle \notag \\
 &\rightarrow & \left(\frac{c^2}{2}+\frac{11}{5} c\right)^{-1/2}\left(:\widetilde{T}(\bar{u})^2:-\frac{3}{10}\partial^2 \widetilde{T}(\bar{u})\right),\nn\\
 \mid 2,\widetilde{2}\rangle&=&\frac{2}{c}L_{-2}\widetilde{L_{-2}} {\mid}0\rangle\rightarrow \frac{2}{c}T(u)\widetilde{T}(\bar{u}).
	\eea
Therefore, the thermal density matrix is
\begin{equation}
	\begin{split}
	\rho&=\frac{e^{-\beta H}}{\Tr(e^{-\beta H})}\\
	&=\frac{|0\rangle\langle0|+\sum |\phi_2\rangle\langle\phi_2|e^{-\frac{4\pi}{TL}}+\sum |\phi_3\rangle\langle\phi_3| e^{-\frac{6\pi}{TL}}+\sum |\phi_4\rangle\langle\phi_4| e^{-\frac{8\pi}{TL}}}{1+e^{-\frac{4\pi}{TL}}+e^{-\frac{6\pi}{TL}}+2e^{-\frac{8\pi}{TL}}}.
	\end{split}
	\end{equation}	
The $n$-th R\'enyi entropy reads
\begin{equation}\label{Sn}	S_n=\frac{1}{1-n}\big{[}\log(\tr(\tr_B|0\rangle\langle0|)^n)+(A_2-2n)e^{-\frac{4\pi}{TL}}+(A_3-2n)e^{-\frac{6\pi}{TL}}+A_4e^{-\frac{8\pi}{TL}}\big{]}
	\end{equation}
The first term in (\ref{Sn}) gives the well-known universal form of R\'enyi entropy for a  single interval on a circle
	\begin{equation}
	\frac{1}{1-n}\log(\tr(\tr_B|0\rangle\langle0|)^n)=\frac{c(1+n)}{12n}\log\sin^2(\frac{\pi l}{L})
	\end{equation}
The coefficients $A_2,A_3$ and $A_4$ are  the contributions from the level 2,3 and 4 excitations respectively. The explicit forms of $A_2$ and $A_3$ have been worked out in \cite{Chen:2014unl}
\begin{equation}
		\begin{split}
		A_2=&2n\frac{\Tr\left[\tr_B|2\rangle\langle2|\left(\tr_B|0\rangle\langle0|\right)^{n-1}\right]}{\Tr\left(\tr_B|0\rangle\langle0|\right)^n}\\
		=&2n\frac{2}{c}\langle w^4 T(w)T(u)\rangle|_{w\rightarrow\infty,u=0}\\
		=&\frac{nc}{9} \frac{(n^2-1)^2}{n^4} \sin^4\frac{\pi l}{L}+\frac{2}{n^3} \frac{\sin^4\frac{\pi l}{L}}{\sin^4\frac{\pi l}{n L}},
		\end{split}
	\end{equation}
\begin{equation}
		\begin{split}
		A_3=&2n\frac{\Tr\left[\tr_B|3\rangle\langle3|\left(\tr_B|0\rangle\langle0|\right)^{n-1}\right]}{\Tr\left(\tr_B|0\rangle\langle0|\right)^n}\\
		&=2n\frac{1}{2c}\langle\left(-w^6\partial T(w)-4w^5T(w)\right)\partial T(u)\rangle|_{w\rightarrow\infty,u=0}\\
		&=\frac{4cn}{9} (1-\frac{1}{n^2})^2 \sin^4\frac{\pi l}{L} \cos^2\frac{\pi l}{L} +\frac{2}{n^5}\bigg{(}5\frac{\sin^6\frac{\pi l}{L}}{\sin^6\frac{\pi l}{nL}}-4\frac{\sin^6\frac{\pi l}{L}}{\sin^4\frac{\pi l}{nL}} \bigg{)}\\
		&-\frac{16}{n^4}\frac{\sin^5\frac{\pi l}{L}}{\sin^5\frac{\pi l}{nL}} \cos\frac{\pi l}{L} \cos\frac{\pi l}{nL}+\frac{8}{n^3}\frac{\sin^4\frac{\pi l}{L}}{\sin^4\frac{\pi l}{nL}} \cos^2\frac{\pi l}{L}.
		\end{split}
	\end{equation}
The computation of $A_4$ is more complicated. It includes several terms 	
	\begin{equation}
	\begin{split}
	A_4=&2n\frac{\Tr\left[\tr_B|4,1\rangle\langle4,1|\left(\tr_B|0\rangle\langle0|\right)^{n-1}\right]}{\Tr\left(\tr_B|0\rangle\langle0|\right)^n}+2n\frac{Tr\left[tr_B|4,2\rangle\langle4,2|\left(tr_B|0\rangle\langle0|\right)^{n-1}\right]}{Tr\left(tr_B|0\rangle\langle0|\right)^n}\\
	&-n^2\left(\frac{\Tr\left[\tr_B|2\rangle\langle2|\left(\tr_B|0\rangle\langle0|\right)^{n-1}\right]}{\Tr\left(\tr_B|0\rangle\langle0|\right)^n}\right)^2\\
	&+n\sum_{j=1}^{n-1}\frac{\Tr\left[\tr_B|2\rangle\langle2|\left(\tr_B|0\rangle\langle0|\right)^{j-1}\tr_B|2\rangle\langle2|\left(\tr_B|0\rangle\langle0|\right)^{n-1-j}\right]}{Tr\left(tr_B|0\rangle\langle0|\right)^n}-3n
	\end{split}
	\end{equation}
The first three terms are the two-point functions on the $n$-sheeted cylinder
with two vertex operators inserting at the infinity future and infinity past on one sheet
\be \frac{\Tr[\tr_B\mid O_1\rangle\langle O_2\mid(\tr_B\mid 0\rangle\langle0\mid)^{(n-1)}]}{\Tr(\tr_B\mid 0\rangle\langle 0\mid)^n}. \ee
The other is the four-point function
\be \frac{\Tr[\tr_B\mid O_1\rangle\langle  O_2\mid(\tr_B\mid 0\rangle\langle 0\mid)^{(j-1)}\tr_B\mid \widetilde{O_1}\rangle\langle \widetilde{O_2}\mid(\tr_B\mid 0\rangle\langle 0\mid)^{(n-j-1)}]}{\Tr(\tr_B\mid 0\rangle\langle  0\mid)^n} \ee
where $O_1, O_2$ are two vertex operators inserting at the infinity future and the infinity past on one sheet, while $ \widetilde{O_1}, \widetilde{O_2}$ are two other vertex operators inserting at the infinity future and the infinite past on another sheet. In \cite{Chen:2014unl}, only the leading contributions linearly proportional to $c$ were obtained. Here we work out all the terms explicitly. The detailed expressions of the terms in $A_4$ are put into the appendix.

Taking all the contributions from $A_2,A_3$ and $A_4$ into account, we finally obtain the $n$-th R\'enyi entropy. It can be separated into two parts: the leading one proportional to $c$, corresponding to the classical part of the holographic R\'enyi entropy, and the next-leading part independent of $c$, corresponding to the 1-loop correction to the holographic R\'enyi entropy. The classical part reads
\bea
S_n^{\mbox{\tiny tree}}&=&\frac{c(1+n)}{n}\left\{\frac{1}{12}\log\sin^2\frac{\pi l}{L}+const\right.\notag \\
&~&-\frac{1}{9}\frac{(n^2-1)}{n^2}\left(\sin^4(\frac{\pi l}{L})
e^{-\frac{4\pi}{TL}}
+4\sin^4(\frac{\pi l}{L})\cos^2(\frac{\pi l}{L})
e^{-\frac{6\pi}{TL}} \right.\notag \\
&~&+\left(\frac{-11-2n^2+1309n^4}{11520n^4}\cos(\frac{8\pi l}{L})-\frac{-11+28n^2+119n^4}{1440n^4}\cos(\frac{6\pi l}{L}) \right.\notag\\
&~&-\frac{77-346n^2+197n^4}{2880n^4}\cos(\frac{4\pi l}{L})-\frac{-77+436n^2+433n^4}{1440n^4}\cos(\frac{2\pi l}{L}) \notag \\
&~&\left. \left. \left. +\frac{-77+466n^2+907n^4}{2304n^4}\right)e^{-\frac{8\pi}{TL}}\right)+O(e^{-\frac{10\pi}{TL}})\right\}.
\label{tree}
\eea
The 1-loop part from the vacuum module is
\begin{equation}\label{Sn21loop}
	\begin{split}
	S_n^{(2)}|_{1-loop}=&\frac{1}{1-n}\left(\frac{2}{n^3} \frac{\sin^4\frac{\pi l}{L}}{\sin^4\frac{\pi l}{n L}}-2n\right)e^{-\frac{4\pi}{TL}}\\
	&+\frac{1}{1-n}\left[\frac{1}{n^3}\frac{\sin^4\frac{\pi l}{L}}{\sin^4\frac{\pi l}{nL}}\left(\frac{2}{n^2}\bigg{(}5\frac{\sin^2\frac{\pi l}{L}}{\sin^2\frac{\pi l}{nL}}-4 \bigg{)}-\frac{8}{n} \sin\frac{2\pi l}{L} \cot\frac{\pi l}{nL}+8 cos^2\frac{\pi l}{L}\right)-2n\right]e^{\frac{-6\pi}{TL}}\\
	&+\frac{1}{1-n}\Bigg\{\frac{1}{n^3}\frac{\sin^4\frac{\pi l}{L}}{\sin^4\frac{\pi l}{nL}}\Bigg[\frac{43}{n^4}\frac{\sin^4\frac{\pi l}{L}}{\sin^4\frac{\pi l}{nL}}+\frac{1}{n^2}\frac{\sin^2\frac{\pi l}{L}}{\sin^2\frac{\pi l}{nL}}\left(140-(140+\frac{60}{n^2})\sin^2\frac{\pi l}{L}-\frac{60}{n}\sin\frac{2\pi l}{L}\cot\frac{\pi l}{nL}\right)\\
	&+\sin^4\frac{\pi l}{L}\left(\frac{260}{9}+\frac{164}{9n^4}+\frac{1016}{9n^2}\right)-\sin^2\frac{\pi l}{L}\left(\frac{152}{3}+\frac{328}{3n^2}\right)+20\\
	&+\frac{1}{n}\sin\frac{2\pi l}{L}\cot\frac{\pi l}{nL}\left[-40+(\frac{136}{3}+\frac{104}{3n^2})\sin^2\frac{\pi l}{L}\right]\Bigg]\\
	& +\frac{\sin^8(\frac{\pi l}{L})}{n^7} \left(\sum_{j=1}^{n-1}\frac{1}{\sin^8\left(\frac{\pi j}{n}\right)}+\sum_{j=1}^{n-1}\frac{16}{\left[\cos\left(\frac{2 \pi l}{nL}\right)-\cos\frac{2 \pi j}{n}\right]^4}\right)-3n\Bigg\}e^{\frac{-8\pi}{TL}}+O(e^{-\frac{10\pi}{TL}})
	\end{split}
	\end{equation}

Taking the $n\to 1$ limit, we find the entanglement entropy from the vacuum module
\begin{equation}
	\begin{split}
	S_{EE}^{(2)}&=\lim_{n\rightarrow 1}S_n^{(2)}\\
	&=c \left[\frac{1}{6} \log(\sin^2\frac{\pi l}{L})+const\right]\\
	&+8\left[1-\frac{\pi l}{L}\cot(\frac{\pi l}{L})\right]e^{-\frac{4 \pi}{TL}}+12\left[1-\frac{\pi l}{L}\cot(\frac{\pi l}{L})\right]e^{-\frac{6 \pi}{TL}}+24\left[1-\frac{\pi l}{L}\cot(\frac{\pi l}{L})\right]e^{-\frac{8 \pi}{TL}}\\
	&+\Bigg\{-\frac{128}{315}\sin^8(\frac{\pi l}{L})+\frac{1}{98304\sin\frac{\pi l}{L}\cos^7\frac{\pi l}{L}}\Big(216\frac{\pi l}{L}+432\frac{\pi l}{L} \cos\frac{2\pi l}{L}+144\frac{\pi l}{L}\cos\frac{4\pi l}{L}+\\
	&48\frac{\pi l}{L}\cos\frac{6\pi l}{L}-108\sin\frac{2\pi l}{L}-84\sin\frac{4\pi l}{L}-44\sin\frac{6\pi l}{L}-3\sin\frac{8\pi l}{L}\Big)\Bigg\}e^{-\frac{8 \pi}{TL}}+O(e^{-\frac{10\pi}{TL}})
	\end{split}
	\end{equation}

Next let us consider the contribution from ${\cal W}$ primary operators. To level 4, there are contributions from ${\cal W}_3$ and ${\cal W}_4$ fields. For the ${\cal W}_3$ field,
the level 3, 4 states are labelled respectively as
\bea
	|3,2\rangle&\rightarrow&\mathcal{W}_3(u), \nn\\
		|4,3\rangle&\rightarrow& i\sqrt{\frac{1}{6}}\partial \mathcal{W}_3(u).
	\eea
Their contributions to the R\'enyi entropy read
\begin{equation}\label{Sn31loop}
		S_n^{(3)}=\frac{1}{1-n}\left[(A_{33}-2n)e^{\frac{-6\pi}{TL}}+(A_{43}-2n)e^{\frac{-8\pi}{TL}}\right]
	\end{equation}
	where:
	\begin{equation}
	\begin{split}
	A_{33}=&2n\frac{\Tr\left[\tr_B|3,2\rangle\langle3,2|\left(\tr_B|0\rangle\langle0|\right)^{n-1}\right]}{\Tr\left(\tr_B|0\rangle\langle0|\right)^n}\\
	=&2n\langle w^6 \mathcal{W}_3(w)\mathcal{W}_3(u)\rangle|_{w\rightarrow\infty,u=0}\\
	=&\frac{2}{n^5}\frac{\sin^6(\frac{\pi l}{L})}{\sin^6(\frac{\pi l}{nL})}
	\end{split}
	\end{equation}
and	
	\begin{equation}
	\begin{split}
	A_{43}=&2n\frac{\Tr\left[\tr_B|4,3\rangle\langle4,3|\left(\tr_B|0\rangle\langle0|\right)^{n-1}\right]}{\Tr\left(\tr_B|0\rangle\langle0|\right)^n}\\
	=&-2n\frac{1}{6}\langle\left( w^8\partial \mathcal{W}_3(w)+6w^7\mathcal{W}_3(w)\right)\partial \mathcal{W}_3(u)\rangle|_{w\rightarrow\infty,u=0}\\
	=&\frac{1}{n^3}\frac{\sin^4(\frac{\pi l}{L})}{\sin^4(\frac{\pi l}{nL})}\left[\frac{14}{n^4}\frac{\sin^4(\frac{\pi l}{L})}{\sin^4(\frac{\pi l}{nL})}+\frac{12}{n^2}\frac{\sin^2(\frac{\pi l}{L})}{\sin^2(\frac{\pi l}{nL})}\left(1-(1+\frac{1}{n^2})\sin^2(\frac{\pi l}{L})-\frac{1}{n}\sin(\frac{2\pi l}{L})\cot(\frac{\pi l}{nL})\right)\right]
	\end{split}
	\end{equation}
The contribution to the entanglement entropy is
\begin{equation}
	\begin{split}
	 S_{EE}^{(3)}&=\lim_{n\rightarrow 1} S_n^{(3)}\\
	&=12\left[1-\frac{\pi l}{L}\cot(\frac{\pi l}{L})\right]e^{-\frac{6 \pi}{TL}}+16\left[1-\frac{\pi l}{L}\cot(\frac{\pi l}{L})\right]e^{-\frac{8 \pi}{TL}}
	\end{split}
	\end{equation}

For the ${\cal W}_4$ field, we only need to consider the level 4 state
\begin{equation}
		|4,4\rangle\rightarrow\mathcal{W}_4(u).
	\end{equation}
	Its contribution to the R\'enyi entropy is:
		\begin{equation}\label{Sn41loop}
	 S_n^{(4)}=\frac{1}{1-n}\left[(A_{44}-2n)e^{\frac{-8\pi}{TL}}\right]
	\end{equation}
	where:
	\begin{equation}
	\begin{split}
	A_{44}=&2n\frac{\Tr\left[\tr_B|4,4\rangle\langle4,4|\left(\tr_B|0\rangle\langle0|\right)^{n-1}\right]}{\Tr\left(\tr_B|0\rangle\langle0|\right)^n}\\
	=&2n\langle w^8 \mathcal{W}_4(w)\mathcal{W}_4(u)\rangle|_{w\rightarrow\infty,u=0}\\
	=&\frac{2}{n^7}\frac{\sin^8(\frac{\pi l}{L})}{\sin^8(\frac{\pi l}{nL})},
	\end{split}
	\end{equation}
	and its contribution to the entanglement entropy is:
	\begin{equation}
	\begin{split}
	 S_{EE}^{(4)}&=\lim_{n\rightarrow 1} S_n^{(4)}\\
	&=16\left[1-\frac{\pi l}{L}\cot(\frac{\pi l}{L})\right]e^{-\frac{8 \pi}{TL}}
	\end{split}
	\end{equation}

Summing all the contributions from the vacuum module and ${\cal W}$ primaries, we find
\be
S_n=S_n^{(2)}+S_n^{(3)}+S_n^{(4)}
\ee
and
\be
S_{EE}=S_{EE}^{(2)}+S_{EE}^{(3)}+S_{EE}^{(4)}.
\ee
Several remarks are in order:
\begin{itemize}
\item The leading contribution in the large central charge limit comes only from the vacuum module, as suggested in \cite{Hartman:2013mia} and confirmed in the two-interval case\cite{Chen:2013kpa,Chen:2013dxa}. The contributions from the ${\cal W}$ fields appear at the subleading order, independent of the central charge. It can be checked easily that the other primary fields contribute at the subleading order as well. Therefore, if the spectrum of the other primary fields is not very dense, the number of the primary fields being not as large as the $e^{\pi c/6}$, their total contributions are only of order 1.
\item The leading contribution of the entanglement entropy takes a universal form, the one for single interval on a circle at zero temperature. The thermal corrections to the entropy originate from the excitations, and appear in the subleading terms. Due to the presence of such correction, the symmetry $l\to L-l$ is broken. In the dual bulk picture, the classical part of the holographic entanglement entropy is symmetric under the transformation $l\to L-l$, since the bulk background is thermal AdS and the geodesic ending on the interval $A$ is the same as the one ending on its complement. However, once we consider the quantum correction, the symmetry is broken.
\item Even though the thermal corrections to the entanglement entropy appear at the subleading order, the thermal corrections to the $n$-th R\'enyi entropy ($n\neq 1$) start to appear at the leading order.
\item In \cite{Cardy:2014jwa}, it was showed that there are universal thermal corrections to the R\'enyi and entanglement entropies:
    \bea
    \delta S_n &=& \frac{g}{1-n}\left(\frac{1}{n^{2\Delta-1}}\frac{\sin^{2\Delta}\left(\frac{\pi l}{L}\right)}{\sin^{2\Delta}\left(\frac{\pi l}{nL}\right)}-n\right)e^{-2\pi \Delta /TL}+o(e^{-2\pi \Delta /TL}) \nn\\
    \delta S_{EE} &=& 2g \Delta \left(1-\frac{\pi l}{L}\cot \left(\frac{\pi l}{L}\right)\right)e^{-2\pi \Delta /TL}+o(e^{-2\pi \Delta /TL}),
    \eea
    where $\Delta$ is the smallest scaling dimension among the set of operators including the stress tensor and all the primaries, and $g$ is their degeneracy. From our computation, we see that the contributions to the R\'enyi entropy from each primary and the stress tensor all take the above universal form, while for the contributions to the entanglement entropy, even the first descendants of the stress tensor and the primaries give the universal thermal corrections. But obviously the thermal corrections from the level 4 states of the vacuum module do not take the above universal form.
\item In the above discussion, the primaries and the stress tensor were treated separately. This may not be true if we extend the study to the higher levels. As shown in \cite{Chen:2013dxa},  the excitations in the ${\cal W}$ primary module may include the combination of the stress tensor and the ${\cal W}$ fields. On the other hand, in the dual bulk side the contributions from the fluctuations with different spin can always be separated.

\end{itemize}

\section{Holographic computation}

From AdS$_3$/CFT$_2$ correspondence, the  R\'enyi entropy of 2D CFT could be computed holographically. As there is a duality between low temperature and high temperature cases,
\be\label{Sdual} L\rightarrow i\beta, \hs{3ex}~\beta\rightarrow iL, \ee
we will discuss the high temperature case, following the discussion in \cite{Barrella:2013wja,Chen:2014unl}

First of all, one has to find the  gravitational configuration ending on the boundary Riemann surface resulted from the replica trick. In the case of single interval on a torus, the boundary Riemann surface is $n$-sheeted torus, which is of genus $n$. If the interval is short, the Schotkky group generating the Riemann surface has been discussed in \cite{Barrella:2013wja}. The uniformization map is determined by the differential equation
\be\label{diff} \psi^{''}(z)+\frac{1}{2}T(z)\psi(z)=0, \ee
where
\be T(z)=\sum_i(\Delta \wp(z-z_i)+\gamma_i\zeta(z-z_i)+\delta) \ee
in the finite temperature case, with $\wp,~\zeta$ being the Weierstrass elliptic function and Weierstrass zeta function respectively, $\g_i$'s being the accessory parameters and $\d$ being an additional constant. The ratio of two independent solutions of (\ref{diff}) defines the Schottky uniformization. The Schottky uniformization could be extended to the bulk to find the gravitational configurations. It turns out that the on-shell regulated action of the gravitational configuration depends on the accessory parameters via the differential equation\cite{Faulkner:2013yia}
\be\label{branch} \frac{\partial S_n^{\gamma}}{\partial z_i}=-\frac{cn}{6(n-1)}\gamma_i. \ee
Moreover, in the $n$-sheeted torus case, one has to take into account of the size dependence of the action in the high temperature case\cite{Chen:2014unl}
\be\label{size} \frac{\partial S_n}{\partial L}=\frac{c}{12\pi}\frac{n}{n-1}\beta(\tilde{\delta}-{\tilde \delta}_{n=1}) \ee
where $\tilde \delta$ is a new constant absorbing all the constant terms in $\wp,~\zeta$.

The accessory parameters $\g_i$ are fixed by imposing the monodromy condition. In the torus case, there are one thermal cycle and one spatial cycle. If the interval is very large, the cycle enclosed the interval should be of trivial monodromy, and one of two cycles of the torus can be imposed the trivial monodromy so that the other cycle is the generator of the Schottky group, on each sheet. In the high temperature, the thermal cycle should be of trivial monodromy. Due to the replica symmetry, there is only one independent accessory parameter, which can be computed perturbatively along with $\tilde \delta$. Integrating the differential equations (\ref{branch}) and (\ref{size}), we obtain the classical part of the holographic R\'enyi entropy, which is in exact agreement with (\ref{tree}) after making duality transformation (\ref{Sdual}).

For the 1-loop quantum correction to the holographic R\'enyi entropy, one may use the heat kernel method to compute, as the gravitational configuration is the quotient of global AdS$_3$ spacetime. After finding the elements of the primitive conjugate class of the Schottky group, the 1-loop partition function reads
\cite{Yin:2007gv,Giombi:2008vd,Barrella:2013wja}
\be\label{1loop} \log Z\mid_{1-loop}=-\sum_{\gamma}\sum_s\sum_{m=s}^{\infty}\log\mid 1-q_{\gamma}^m\mid, \ee
where $\gamma$'s are the primitive words, and $q_{\gamma}^{-\frac{1}{2}}$ is the  eigenvalue of $\gamma$, $s$ is the spin of the fluctuations. Obviously, the contributions from different spin $s$ fluctuations can be computed separately.
 Though there are infinite number of the primitive words, only finite number of them contribute to the partition function up to a fixed order of $e^{-\pi TL}$. To our purpose, we would like to compute the quantum correction up to order $e^{-8\pi TL}$. In the following, we introduce two parameters defined by
 \begin{equation}
	u_R=e^{-2\pi TL} \qquad u_y=e^{-2\pi Ty}
	\end{equation}

For the spin $2$ fluctuations, the 1-loop partition function becomes
\begin{equation}
	\log Z^{(2)}_n|_{1-loop}=(2nq_1^2+2nq_1^3+3nq_1^4)+\sum_{\gamma_2}q_{\gamma_2}^2,
	\label{Z_n}
	\end{equation}
	where $q_1^{-1/2}$ is the eigenvalue of $L_i$ and $L_i^-$, and $q_{\gamma_2}^{-1/2}$ is the eigenvalue of $L_iL_j$, $L_i^-L_j^-$ and $L_iL_j^-$. To read the contribution up to order $u_R^4$, we need the expansion of $q_1$ to the order $u_R^2$.

For the generator
\begin{equation}
	L_{k_1}^{\sigma_1}L_{k_2}^{\sigma_2}...L_{k_m}^{\sigma_m},
	\end{equation}
	with $\sigma_j=\pm 1$, $j=1,2,...m$, the eigenvalue is\cite{Barrella:2013wja}
\begin{equation}
	q_{\gamma_2}^{-1/2}=\left[\frac{nu_y}{(1-u_y^2)\sqrt{u_R}}\right]^m \prod_{j=1}^{m}\sigma_j(u_y^{-\sigma_j/n}-e^{2\pi i(k_j-k_{j+1})/n}u_y^{\sigma_{j+1}/n})+\mathcal{O}(u_R^{-m/2+1}).
	\end{equation}
To read the contributions to the order $u_R^4$, we only need the leading term in $q_{\gamma_2}^{-1/2}$. Therefore, for the generators $L_{k_1}L_{k_2}$ and $L_{k_1}^-L_{k_2}^-$, we have
	\begin{equation}
	q_{\gamma_2}^2=16 \frac{\sinh^8(2 \pi Ty)}{n^8} u_R^4\left[\cosh\left(\frac{4 \pi Ty}{n}\right)-\cosh\frac{2 \pi (k_1-k_2)}{n}\right]^{-4}.
	\end{equation}
		For the generators $L_{k_1}L_{k_2}^-$, we find
	\begin{equation}
	q_{\gamma_2}^2= \frac{\sinh^8(2 \pi Ty)}{n^8} u_R^4\frac{1}{\sinh^8\left(\frac{\pi (k_1-k_2)}{n}\right)}.
	\end{equation}
 Finally we obtain
 \begin{equation}
	\sum_{\gamma_2}q_{\gamma_2}^2=\frac{\sinh^8(2 \pi Ty)}{n^7} u_R^4\left[\sum_{j=1}^{n-1}\frac{1}{\sinh^8\left(\frac{\pi j}{n}\right)}+\sum_{j=1}^{n-1}\frac{16}{\left[\cosh\left(\frac{4 \pi Ty}{n}\right)-\cosh\frac{2 \pi j}{n}\right]^4}\right].
	\end{equation}

 On the other hand, to read the eigenvalue $q_1$ up to order $u_R^2$, we need to extend the analysis in \cite{Barrella:2013wja} to higher order. For the generators $L_j$ and $L_j^-$, $q_1$ can be expanded as:
	\begin{equation}
	q_1=a u_R(1+b u_R+c u_R^2)
	\end{equation}
where $a$ and $b$ have been computed before, but $c$ is unknown. The
monodromy along the spatial cycle is:
		\begin{equation}
	L_1=\begin{pmatrix}
	(L_1)_{11} &(L_1)_{12}\\(L_1)_{21} &(L_1)_{22}
	\end{pmatrix}
	\end{equation}
	and $q_1^{-1/2}$ is the larger eigenvalue of $L_1$:
	\begin{equation}
	q_1^{-1/2}=\frac{1}{2}\left[(L_1)_{11}+(L_1)_{22}+\sqrt{((L_1)_{11}+(L_1)_{22})^2-4}\right]
	\end{equation}
	From the action of $L_1$
	\begin{equation}
	\begin{pmatrix}
	\psi^+(u/u_R)\\\psi^-(u/u_R)
	\end{pmatrix}
	=L_1\begin{pmatrix}
	\psi^+(u)\\\psi^-(u)
	\end{pmatrix}
	\end{equation}
	where
	\begin{equation}
	\begin{split}
	\psi^+(u)=&\frac{1}{\sqrt{u}}(u-u_y)^{\Delta_+}\left(u-\frac{1}{u_y}\right)^{\Delta_-} \bigg\{1+\frac{(n^2-1)(u_y^2-1)^2u_R^2}{24 n^3 u^2 u_y^3}\times\\
	&\left[u\left((n+1)u^2+n-1\right)u_y^2+u\left((n-1)u^2+n+1\right)-n(u^4+1)u_y\right]+\mathcal{O}(u_R^3)\bigg\}
	\end{split}
	\end{equation}
	and $\psi^-$ is obtained by substitute $u$ with $1/u$ in $\psi^+$:
\be
\psi^-(u)=\psi^+(1/u),
\ee
	we can find the expansions of the elements in $L_1$:
		\begin{equation}
	\begin{split}
	(L_1)_{11}=&\frac{nu_{y}^{1-\frac{1}{n}}}{(1-u_y^2)\sqrt{u_R}}\bigg{\{}1-\frac{[(n+1)u_y^2+n-1]^2}{4n^2u_y^2}u_R-\frac{1}{3}(1-\frac{1}{n^2})\sinh^2(2\pi Ty)\times\\
	&\times\left[\left(\frac{10}{3}+\frac{2}{3}\frac{1}{n^2}\right)\sinh^2(2\pi Ty)+4-\frac{2}{n}\sinh(4\pi Ty)\right]u_R^2+\mathcal{O}(u_R^3)\bigg{\}}
	\end{split}
	\end{equation}
	
	\begin{equation}
	\begin{split}
	(L_1)_{12}=&\frac{nu_{y}}{(1-u_y^2)\sqrt{u_R}}\bigg{\{}1-\frac{[(n+1)u_y^2+n-1][(n-1)u_y^2+n+1]}{4n^2u_y^2}u_R\\
	&-\frac{1}{3}(1-\frac{1}{n^2})\sinh^2(2\pi Ty)\left[\left(\frac{10}{3}+\frac{2}{3}\frac{1}{n^2}\right)\sinh^2(2\pi Ty)+4\right]u_R^2+\mathcal{O}(u_R^3)\bigg{\}}
	\end{split}
	\end{equation}
	
	\begin{equation}
	(L_1)_{21}=-(L_1)_{12}
	\end{equation}
	
	\begin{equation}
	(L_1)_{22}=(L_1)_{11}|_{n\rightarrow-n}
	\end{equation}
	
The other Schottky generators are obtained by
	\begin{equation}
	L_i=M_2^{i-1}L_1M_2^{-(i-1)}
	\end{equation}
	where
	
	\begin{equation}
	M_2=\begin{pmatrix}
	e^{2\pi i\Delta_+} &0\\0 &e^{2\pi i \Delta_{-}}
	\end{pmatrix}
	\end{equation}
	For all the generators $L_i$ and their inverse, their eigenvalue is $q_1^{-1/2}$. The coefficients in $q_1$ can be read straightforwardly
		\begin{equation}
	a=\frac{1}{n^2}\frac{\sinh^2 2\pi Ty}{\sinh^2\frac{2\pi Ty}{n}}
	\end{equation}
	\begin{equation}
	b=\frac{2}{n^2}\frac{\sinh^2 2\pi Ty}{\sinh^2\frac{2\pi Ty}{n}}-\frac{2}{n} \sinh(4\pi Ty) \coth(\frac{2\pi Ty}{n})+2+2(1+\frac{1}{n^2})\sinh^2(2\pi Ty)
	\end{equation}
	
	\begin{equation}
	\begin{split}
	c=&\frac{5}{n^4}\frac{\sinh^4 2\pi Ty}{\sinh^4\frac{2\pi Ty}{n}}+\frac{1}{n^2}\frac{\sinh^2 2\pi Ty}{\sinh^2\frac{2\pi Ty}{n}}\left[20+\left(20+\frac{8}{n^2}\right)\sinh^2(2\pi Ty)-\frac{8}{n}\sinh(4\pi Ty)\coth(\frac{2\pi Ty}{n})\right]\\
	&-\frac{1}{n}\sinh(4\pi Ty)\coth(\frac{2\pi Ty}{n})\left[6+\left(\frac{22}{3}+\frac{14}{3}\frac{1}{n^2}\right)\sinh^2(2\pi Ty)\right]+3\\
	&+\left(\frac{26}{3}+\frac{46}{3}\frac{1}{n^2}\right)\sinh^2(2\pi Ty)+\left(\frac{47}{9}+\frac{23}{9n^4}+\frac{146}{9n^2}\right)\sinh^4(2\pi Ty)
	\end{split}
	\end{equation}
	
After taking into account of all the contributions from the generators, we 	get:
	\bea
	\lefteqn{\log Z_n^{(2)}|_{1-loop}=(2nq_1^2+2nq_1^3+3nq_1^4)+\sum_{\gamma_2}q_{\gamma_2}^2} \nn \\
	&=&\frac{1}{n^3} \frac{\sinh^42\pi Ty}{\sinh^4\frac{2\pi Ty}{n}}\Bigg\{2u_R^2
\nn\\
&&+\frac{\sinh^42\pi Ty}{\sinh^4\frac{2\pi Ty}{n}}\left(\frac{2}{n^2}\bigg{(}5\frac{\sinh^22\pi Ty}{\sinh^2\frac{2\pi Ty}{n}}+4\sinh^22\pi Ty \bigg{)}-\frac{8}{n} \sinh(4\pi Ty) \coth\frac{2\pi Ty}{n}+8 \cosh^22\pi Ty\right)u_R^3\nn\\
	&&+\Bigg[\frac{43}{n^4}\frac{\sinh^42\pi Ty}{\sinh^4\frac{2\pi Ty}{n}}+\frac{1}{n^2}\frac{\sinh^22\pi Ty}{\sinh^2\frac{2\pi Ty}{n}}\left(140+(140+\frac{60}{n^2})\sinh^22\pi Ty-\frac{60}{n}\sinh(4\pi Ty)\coth\frac{2\pi Ty}{n}\right)\nn\\
	&&+\sinh^42\pi Ty\left(\frac{260}{9}+\frac{164}{9n^4}+\frac{1016}{9n^2}\right)+\sinh^22\pi Ty\left(\frac{152}{3}+\frac{328}{3n^2}\right)+20\nn\\
	&&+\frac{1}{n}\sinh(4\pi Ty)\coth\frac{2\pi Ty}{n}\left[-40-(\frac{136}{3}+\frac{104}{3n^2})\sinh^22\pi Ty\right]\nn\\
	&& +\frac{\sinh^4\frac{2\pi Ty}{n}\sinh^4 2\pi Ty}{n^4} \left(\sum_{j=1}^{n-1}\frac{1}{\sinh^8\left(\frac{\pi j}{n}\right)}+\sum_{j=1}^{n-1}\frac{16}{\left[\cosh\left(\frac{2 \pi Ty}{n}\right)-\cosh\frac{2 \pi j}{n}\right]^4}\right)\Bigg]u_R^4+O(u_R^5)\Bigg\}.
		\eea

Similarly we find the quantum correction from spin 3 and spin 4 fluctuations. For the spin 3 fluctuation, we have
\begin{equation}
		\begin{split}
		 \log Z_n^{(3)}|_{1-loop}=&2nq_1^3+2nq_1^4\\
		=&\frac{2}{n^5}\frac{\sinh^6(2\pi Ty)}{\sinh^6(\frac{2\pi Ty}{n})}u_R^3+\frac{1}{n^3}\frac{\sinh^4(2\pi Ty)}{\sinh^4(\frac{2\pi Ty}{n})}\Bigg[\frac{14}{n^4}\frac{\sinh^4(2\pi Ty)}{\sinh^4(\frac{2\pi Ty }{n})}\\
		&+\frac{12}{n^2}\frac{\sinh^2(2\pi Ty)}{\sinh^2(\frac{2\pi Ty}{n})}\left(1+(1+\frac{1}{n^2})\sinh^2(2\pi Ty)-\frac{1}{n}\sinh(4\pi Ty)\coth(\frac{2\pi Ty}{n})\right)\Bigg]u_R^4\\
&+O(u_R^5).
		\end{split}
	\end{equation}
 For the spin 4 fluctuations, we have
\begin{equation}
	\begin{split}
	 \log Z_n^{(4)}|_{1-loop}=&2nq_1^4\\
	=&\frac{2}{n^7}\frac{\sinh^8(2\pi Ty)}{\sinh^8(\frac{2\pi Ty}{n})}u_R^4
	\end{split}
	\end{equation}

The 1-loop quantum correction to the holographic R\'enyi entropy is
\be
S_n|_{1-loop}=\frac{1}{1-n}(\log Z_n - n\log Z_1).
\ee
After making the modular transformation (\ref{Sdual}), we find the exact agreement with the field theory results $S_n^{(2)}|_{1-loop}, S_n^{(3)}|_{1-loop}, S_n^{(4)}|_{1-loop}$ in (\ref{Sn21loop})(\ref{Sn31loop})(\ref{Sn41loop}) respectively.

\section{Conclusion and Discussion}

In this paper, we extend our study of the holographic R\'enyi entropy of single interval on a torus to the CFT with ${\cal W}$ symmetry. To read the possible correction from $W$ fields and higher spin fluctuations, we computed the thermal correction to order $e^{-8\pi /TL}$. In the field theory, this requires us to compute the contribution from all the level 4 states, from both the vacuum module and $W$ primaries. On the gravity side, we need to compute the eigenvalue of single Schottky generators to the order $e^{-4 \pi/TL}$. We found complete agreement between the  bulk and boundary theories.

In this work, we focused on the case that the interval is short such that we may apply the strategy in \cite{Chen:2014unl} to do computation. When the interval is very large $L-l\sim o(\epsilon)$, it turns out that both the field theory and gravity computations should be reconsidered. On the field theory side, we need a different expansion formalism. On the bulk side, we need to find another set of monodromy condition. In \cite{Chen:2015kua}, the large interval limit of holographic R\'enyi entropy has been studied in pure AdS$_3$ gravity. It would be interesting to generalize the study to the case with a higher spin symmetry.

The study in this paper support the picture that the classical part of the holographic R\'enyi entropy is captured by the vacuum module of the CFT in the large central charge limit, and the quantum part should be reproduced by considering all the modules. In particular, the precise agreement beyond classical level suggest that the correspondence must be true even at quantum level. On the 1-loop level, we have a universal form (\ref{1loop}). It would be interesting to prove the 1-loop agreement for general cases\cite{ChenWu}.

\vspace*{10mm}
\noindent {\large{\bf Acknowledgments}}\\
The work was  supported in part by NSFC Grants No.~11275010, No.~11335012 and No.~11325522.
ZC was supported by the Yu-gang Mao funding for undergraduates. 
\vspace*{5mm}

\section*{Appendix: The coefficients of $A_4$}

The contributions from the level 4 states of the vacuum module is given by $A_4$:
		\begin{equation}
	\begin{split}
	A_4=&2n\frac{\Tr\left[\tr_B|4,1\rangle\langle4,1|\left(\tr_B|0\rangle\langle0|\right)^{n-1}\right]}{\Tr\left(\tr_B|0\rangle\langle0|\right)^n}
+2n\frac{\Tr\left[\tr_B|4,2\rangle\langle4,2|\left(\tr_B|0\rangle\langle0|\right)^{n-1}\right]}{\Tr\left(\tr_B|0\rangle\langle0|\right)^n}\\
	&-n^2\left(\frac{\Tr\left[\tr_B|2\rangle\langle2|\left(\tr_B|0\rangle\langle0|\right)^{n-1}\right]}{\Tr\left(\tr_B|0\rangle\langle0|\right)^n}\right)^2\\
	&+n\sum_{j=1}^{n-1}\frac{\Tr\left[\tr_B|2\rangle\langle2|\left(\tr_B|0\rangle\langle0|\right)^{j-1}\tr_B|2\rangle\langle2|\left(\tr_B|0\rangle\langle0|\right)^{n-1-j}\right]}{\Tr\left(\tr_B|0\rangle\langle0|\right)^n}-3n
	\end{split}
	\end{equation}
The first term is a two-point function, given by
	\begin{equation}
	\begin{split}
	&2n\frac{\Tr\left[\tr_B|4,1\rangle\langle4,1|\left(\tr_B|0\rangle\langle0|\right)^{n-1}\right]}{\Tr\left(\tr_B|0\rangle\langle0|\right)^n}\\
	=&\frac{2n}{5c}\langle\big(\frac{1}{2}w^8\partial^2T(w)+5w^7\partial T(w)+10w^6 T(w)\big)\frac{1}{2}\partial^2T(u)\rangle|_{w\rightarrow \infty,u=0}\\
	=&\frac{2nc}{45}(1-\frac{1}{n^2})^2\sin^4(\frac{\pi l}{L})(3\cos\left(\frac{2\pi l}{L}\right)+2)^2+\frac{1}{n^3}\frac{\sin^4(\frac{\pi l}{L})}{\sin^4(\frac{\pi l}{nL})}\Bigg\{42\frac{1}{n^4}\frac{\sin^4(\frac{\pi l}{L})}{\sin^4(\frac{\pi l}{nL})}\\
	&+\frac{1}{n^2}\frac{\sin^2(\frac{\pi l}{L})}{\sin^2(\frac{\pi l}{nL})}\Big(140-(148+\frac{52}{n^2})\sin^2(\frac{\pi l}{L})-\frac{60}{n}\sin(\frac{2\pi l}{L})\cot(\frac{\pi l}{nL})\Big)+20\\
	& +\sin^2(\frac{\pi l}{L})(-48-\frac{112}{n^2})+\sin^4(\frac{\pi l}{L})(\frac{144}{5}+\frac{64}{5n^4}+\frac{592}{5n^2})\\
	&+\frac{1}{n}\sin(\frac{2\pi l}{L}) \cot(\frac{\pi l}{nL})\big((\frac{32}{n^2}+48)\sin^2(\frac{\pi l}{L})-40\big)\Bigg\}
	\end{split}
	\end{equation}
	
	For the second term, we have
	\begin{equation}
	\begin{split}
	& 2n\frac{\Tr\left[\tr_B|4,2\rangle\langle4,2|\left(\tr_B|0\rangle\langle0|\right)^{n-1}\right]}{\Tr\left(\tr_B|0\rangle\langle0|\right)^n}\\
	=&2n(\frac{1}{2}c^2+\frac{11}{5}c)^{-1}\langle w^8\big(:T(w)^2:-\frac{3}{10}\partial^2T(w)\big)\big(:T(u)^2:-\frac{3}{10}\partial^2 T(u)\big)\rangle|_{w\rightarrow\infty,u=0}\\
	=&\frac{c^2 n}{324}(1-\frac{1}{n^2})^4\sin^8(\frac{\pi l}{L})+2n \frac{11c}{1620}(1-\frac{1}{n^4})^4\sin^8(\frac{\pi l}{L})+2n \frac{c}{9}(1-\frac{1}{n^4})^2\frac{\sin^8(\frac{\pi l}{L})}{\sin^4(\frac{\pi l}{nL})}\\
	&+\frac{1}{n^3}\frac{\sin^4(\frac{\pi l}{L})}{\sin^4(\frac{\pi l}{nL})}\left[\frac{2}{n^4}\frac{\sin^4(\frac{\pi l}{L})}{\sin^4(\frac{\pi l}{nL})}+\frac{44}{45} \sin^4(\frac{\pi l}{L})(1-\frac{1}{n^2})^2\right]+\mathcal{O}(c^{-1})
	\end{split}
	\end{equation}
The third term is proportional to the square of $A_2$. 	
	The forth term is a four-point function, and is given by
	\begin{equation}
		\begin{split}
		&n\sum_{j=1}^{n-1}\frac{\Tr\left[\tr_B|2\rangle\langle2|\left(\tr_B|0\rangle\langle0|\right)^{j-1}\tr_B|2\rangle\langle2|\left(\tr_B|0\rangle\langle0|\right)^{n-1-j}\right]}{\Tr\left(\tr_B|0\rangle\langle0|\right)^n}\\
		=&n\sum_{j=2}^{n}(\frac{2}{c})^2\langle w^4T^{(1)}(w)T^{(1)}(u)\widetilde{w}^4T^{(j)}(\widetilde{w})T^{(j)}(\widetilde{u})\rangle\\
		=&n\frac{c^2}{324}(n-1)(1-\frac{1}{n^2})^4\sin^8(\frac{\pi l}{L})+\frac{c}{9}(n-2)(1-\frac{1}{n^2})^2\frac{1}{n^3}\frac{\sin^4(\frac{\pi l}{L})}{\sin^4(\frac{\pi l}{nL})}\\
		&+n\frac{c}{405}(1-\frac{1}{n^2})^2\frac{(n^2+11)(n^2-1)}{n^4}\sin^8(\frac{\pi l}{L})+n\frac{c}{27}(1-\frac{1}{n^2})^2\sin^4(\frac{\pi l}{L})\left[(1-\frac{1}{n^2})\cos(\frac{2\pi l}{L})+(2+\frac{1}{n^2})\right]\\
		&+\frac{1}{n^3}\frac{\sin^4(\frac{\pi l}{L})}{\sin^4(\frac{\pi l}{nL})}\Bigg\{\frac{n-1}{n^4}\frac{\sin^4(\frac{\pi l}{L})}{\sin^4(\frac{\pi l}{nL})}-\frac{8}{3}(1-\frac{1}{n^2})\sin^2(\frac{\pi l}{L})\frac{1}{n}\cot(\frac{\pi l}{nL})\sin(\frac{2\pi l}{L})-\frac{8}{3}(1-\frac{1}{n^2})\sin^2(\frac{\pi l}{L})\\
		&-\frac{8}{9}\sin^4(\frac{\pi l}{L})(1-\frac{1}{n^2})(1+\frac{5}{n^2})+8\frac{1}{n^2}\frac{\sin^2(\frac{\pi l}{L})}{\sin^2(\frac{\pi l}{nL})}(1-\frac{1}{n^2})\sin^2(\frac{\pi l}{L})
		\Bigg\}\\
		&+\frac{\sin^8(\frac{\pi l}{L})}{n^7} \left[\sum_{j=1}^{n-1}\frac{1}{\sin^8\left(\frac{\pi j}{n}\right)}+\sum_{j=1}^{n-1}\frac{16}{\left[\cos\left(\frac{2 \pi l}{nL}\right)-\cos\frac{2 \pi j}{n}\right]^4}\right]
		\end{split}
	\end{equation}

\end{document}